\pdfoutput=1
\documentclass[twoside,twocolumn]{article}
\usepackage[english]{babel}
\usepackage[latin1]{inputenc}
\usepackage{amsmath,amsfonts,bbm}
\usepackage[super,sort&compress,comma]{natbib}
\usepackage{mhchem}
\usepackage{times,mathptmx}
\usepackage{sectsty}
\usepackage{balance}

\usepackage{graphicx}
\usepackage{lastpage}
\usepackage[format=plain,singlelinecheck=false,font=small,labelfont=bf,labelsep=space]{caption}
\usepackage{fancyhdr}
\makeatletter
\def\subsubsection{\@startsection{subsubsection}{3}{10pt}{-1.25ex plus -1ex minus -.1ex}{-1em plus -0.5em}{\normalsize\bf}}
\def\paragraph{\@startsection{paragraph}{4}{10pt}{-1.25ex plus -1ex minus -.1ex}{-1em plus -0.5em}{\normalsize\textit}}
\renewcommand\@biblabel[1]{#1}
\renewcommand\@makefntext[1]{\noindent\makebox[0pt][r]{\@thefnmark\,}#1}
\makeatother

\fancypagestyle{plain}{
  \fancyhead[L]{}
  \fancyhead[C]{}
  \fancyhead[R]{}
  }

\sectionfont{\large}
\subsectionfont{\normalsize}

\newcommand{\bfh}{{\mathbf{h}}}
\newcommand{\bfk}{{\mathbf{k}}}
\newcommand{\bfN}{{\mathbf{N}}}
\newcommand{\bfq}{{\mathbf{q}}}
\newcommand{\bfr}{{\mathbf{r}}}

\newcommand{\bfu}{{\mathbf{u}}}

\newcommand{\bfzero}{{\mathbf{0}}}

\newcommand{\dotp}{\boldsymbol{\cdot}}

\newcommand{\grad}{\operatorname{grad}}
\renewcommand{\div}{\operatorname{div}}
\newcommand{\laplace}{\operatorname{\Delta}}

\newcommand{\curl}{\operatorname{\bf curl}}

\newcommand{\bfnabla}{\boldsymbol{\nabla}}
\newcommand{\bfgrot}{\raisebox{-0.25ex}{\scalebox{0.8}[1.2]{\rotatebox[origin=c]{90}{\reflectbox{$\bfnabla$}}}}}

\newcommand{\Reals}{\mathbbm{R}}

\def\grad{\operatorname{grad}}
\def\div{\operatorname{div}}
\def\sinc{\operatorname{sinc}}
\def\erfi{\operatorname{erfi}}

\def\scoeff{\lambda}
\begin{document}
\thispagestyle{plain}

\twocolumn[\begin{@twocolumnfalse}
  \noindent\LARGE{\textbf{Truncated correlations in video microscopy of colloidal solids}}
  \vspace{0.6cm}

  \noindent\large{\textbf{%
    Michael Schindler,$^{\ast}$\textit{$^{a}$}
    A.C. Maggs\textit{$^{a}$}}}
  \vspace{0.5cm}

  \noindent\textit{\small{\textbf{%
    Received Xth XXXXXXXXXX 20XX, Accepted Xth XXXXXXXXX 20XX\newline
    First published on the web Xth XXXXXXXXXX 20XX}}}

  \noindent \textbf{\small{DOI: 10.1039/b000000x}}
  \vspace{0.6cm}

  \noindent\normalsize{%
    Studies by video microscopy on fluctuating colloids measure the
    real-space cross-correlations in particle motion. This set of
    correlations is then treated as a matrix, in order to study the
    spectrum and mode structure. We show that in general the modes
    are modified by the truncation of the full real-space
    correlations. We perform a theoretical analysis of the truncation,
    find the boundary conditions imposed by the truncation, and
    propose practical windowing strategies to eliminate artefacts. We
    study the problem from various perspectives, to compile a 
    survey for experimentalists.  }\vspace{0.5cm}
\end{@twocolumnfalse}]
\footnotetext{\textit{$^{a}$~Laboratoire~PCT, Gulliver CNRS-ESPCI UMR~7083, 10~rue~Vauquelin, 75231 Paris Cedex 05.}}


\section{Introduction}

Many experiments have studied the fluctuations of two and three
dimensional colloidal solids (both crystalline, amorphous and glassy)
using video and confocal microscopy \cite{antinaPRL, kurchan,ghosh,
  science, maret, claire, maret2, andrea, maret3}. A section of a
large sample is observed and recorded; fluctuations of the subsystem
are then analysed off-line. In particular one wishes to measures the
mode structure and dispersion relations of the medium to extract
material properties such as elastic constants. An important technical
question which comes up is how to perform analysis on data which is
gathered within a window which is smaller than the full sample. One of
the main reasons for working with such truncated data is of course the
hope of eliminating the uncontrolled influence of walls by imaging
well within the sample using confocal techniques. We will show,
however, that truncating the full set of fluctuations outside of the
observation window introduces effective boundary conditions which
rather unexpectedly lead to errors in measurements of material
properties.

The crucial question is how the result of the observed mode structure
converges when the observation volume increases. It is generally
assumed that convergence is assured for large system size. After all,
Weyl's theorem on the density of states of a vibrating system shows
that the density of states is asymptotically independent of the system
shape \cite{weyl}; thus theorists freely interchange the true
boundary conditions of a physical system by a mathematically
convenient choice of periodic modes.  However, we show in this paper
that the assumption is dangerous when using Fourier analysis of
observed amplitudes.  To be concrete let us now summarize the kinds of
data and analysis which are available:

Video or confocal observation of colloidal solids generates data sets consisting
of the position of particles recorded over many frames.  One then
calculates the mean position of each particle and the correlations in
the displacements, $u_i(\bfr)$, by evaluating the two-particle
function
\begin{equation}
  \label{eq:corr}
  C_{ij}(\bfr,\bfr')= \langle u_i(\bfr) u_j(\bfr') \rangle
\end{equation}
with directional indexes $i$, $j$, and with $\bfr$~being the reference
positions of the $N$~particles. The average is over the recorded
frames. We work in $d$-dimensional space where the interesting values
are $d=2, 3$ when imaging colloids at interfaces, or within volumes.

Three methods of analysis suggest themselves:
\begin{enumerate}%
\renewcommand{\theenumi}{(\Alph{enumi})}%
\renewcommand{\labelenumi}{\theenumi}%
\item\label{meth:direct} diagonalisation of the correlation matrix
  eq.~\eqref{eq:corr} of dimension $(d\,N)\times(d\,N)$ in order to
  study its eigenvalues and eigenvectors,
\item\label{meth:recip} evaluation of amplitudes in Fourier
  space\footnote{We differentiate in notation between the reciprocal
  vectors of the full (possibly infinite) system, $\bfk$, and
  $\bfq$~for those of the windowed system.}
  \begin{equation}
    u_i(\bfq) = \sum_{\bfr} e^{i \bfq \dotp \bfr} u_i (\bfr),
  \end{equation}
  followed by a similar study of the matrix
  \begin{equation}
    C_{ij}(\bfq,\bfq')= \bigl\langle u_i{(\bfq)} u_j(\bfq')\bigr\rangle,
  \end{equation}
\item\label{meth:diag} direct use of only the diagonal Fourier
  coefficients
  \begin{equation}
    C_{ij}(\bfq,-\bfq) = \bigl\langle u_i{(\bfq)} u_j{(-\bfq)} \bigr\rangle
  \end{equation}
  which reduces, for a Bravais lattice, to a set of $N$ matrices of
  size $d \times d$.
\end{enumerate}

The full diagonalisation of the matrix $C_{ij}$ allows one to plot
{\sl visually seductive} pictures of mode structure in the truncated
system.  The question arises as to the {\sl exact} link between these
plots and the standard treatment of modes in an elastic medium via the
wave equation. In particular are there effective boundary conditions
introduced by the truncation which influence the modes? In this paper
we show that in a simple square geometry eigenmodes of the correlation
matrix do not have a pure longitudinal or transverse nature, unlike a
bulk sample. We quantify this effect in a study of a two-dimensional
elastic medium. We also note that the decomposition into
longitudinal and transverse is in ambiguous in finite geometries. 

The second formulation in terms of a matrix of Fourier coefficients
requires some care as to the definition of the eigenvalue problem in
order to be equivalent to the real space form \cite{ortho}, in
particular in a disordered system plane waves do not form an
orthonormal basis; thus one must study a generalized, pencil
eigenvalue problem. We will not consider this case further in the
present paper.

The use of the third method, studying $C_{ij}(\bfq,-\bfq)$ seems
particularly practical in large experimental systems because it avoids
the expensive diagonalisation that is required for the other two
cases. It might be expected to give exact results for elastic moduli
in crystals, and good estimates in disordered, non-glassy solids.
However, we will show numerically that $C_{ij}(\bfq,-\bfq)$ is
contaminated by the truncation of the full correlation functions. We
explain the result visually with an analogy with the diffraction
pattern of the observation window. We propose a simple practical
solution to the contamination by the use of alternative windowing
functions with superior properties in Fourier space.

\section{Elastic theory}
In this section we resume the results from elasticity theory that we
will require. In a three-dimensional elastic
medium the quadratic fluctuations about the energy minimum, as well as
the propagating modes of a crystal are deduced from the Christoffel
matrix~\cite{Wallace70}. In a cubic solid this has the form
\begin{equation}
  \label{eq:christoffel}
  D_{ik}(\bfk) = \Bigl[\lambda\delta_{ij}\delta_{kl}
  + \mu(\delta_{ik}\delta_{jl} + \delta_{il}\delta_{jk})
  + \nu S_{ijkl}\Bigr] k_j k_l,
\end{equation}
with Lam\'e constants $\lambda$, $\mu$ and anisotropy $\nu$. This
expression is also valid under uniform, isotropic stress such as the
pressure which must be applied in non-bound colloidal solids.  The
free energy of small fluctuations about the equilibrium position is
then given by the functional
\begin{equation}
  E[\bfu] = \sum_{i,j,\bfk} \frac{1}{2} u_i(\bfk) u_j(-\bfk) D_{ij} (\bfk).
\end{equation}
The Green function of the static elastic problem is then the inverse
of the Christoffel matrix,
\begin{equation}
  \label{eq:inverse}
  D_{ij}(\bfk) G_{jk}(\bfk) = \delta_{ik}.
\end{equation}
It describes the response of the medium to static forces, as well as
correlations in position fluctuations which can be measured in
microscopy. In a face-centred cubic crystal with nearest-neighbour
central potentials $\mu=\lambda=-\nu$, see Eq.~(12.7) of
Ref.~\citenum{Born98}. Face-centred hard-sphere systems and real
experiments have non-linearities that slightly modify the relation
between these three constants \cite{frenkel}.

A similar mathematical structure describes fluctuations in a
two-dimensional hexagonal crystal, with however $\nu=0$, implying that
the long-wavelength mode structure is isotropic with just two types of
modes- longitudinal and transverse. This theory can be used to study
the statistics of colloidal crystals at interfaces \cite{maret}.

For many of the discussions in this paper the most important
characteristic of the matrix eq.~\eqref{eq:christoffel} is the scaling
in~$k^2$. This motivates the study of a scalar energy function
\begin{equation}
  \label{eq:scalar}
  E[u] = \frac{\scoeff}{2} \int (\nabla u)^2\; {\rm d}\bfr
       = \frac{\scoeff}{2} \sum_{\bfk} k^2 |u(\bfk)|^2.
\end{equation}
Such energy functions eliminate the need for detailed tensor analysis
and allow one to transpose well known theorems in potential theory to
our study of truncation artefacts. In particular for eq.~(\ref{eq:scalar}) we find
\begin{equation}
  G(\bfk)= \frac{1}{\scoeff k^2}. \label{eq:scalarG}
\end{equation}
Then the real-space Green (in $d=3$) function is given by the Fourier transform
of eq.~(\ref{eq:scalarG})
\begin{equation}
  G(\bfr-\bfr') = \int e^{i\bfk\dotp(\bfr-\bfr')} G(\bfk)\;
  \frac{d \bfk}{(2\pi)^d} = \frac{1}{4\pi\scoeff |\bfr-\bfr'|}. \label{eq:coulomb}
\end{equation}
The correlations are translationally invariant. The corresponding
correlations for isotropic elasticity can be found in standard references
\cite{landau}, again the decay in separation varies as
$1/|\bfr-\bfr'|$, with additional tensor structure. Expressions with
cubic anisotropy are treated in the recent literature
\cite{Morawiec94}. The experimental correlation matrices are related
via equipartition to the Green function:
\begin{equation}
  \label{eq:equiv}
  \langle u_i(\bfk) u_j(-\bfk) \rangle = k_BT G_{ij}(\bfk),
\end{equation}
where the correlation is is evaluated for the full system and not
truncated to a window.

Experimentalists also study \emph{projected} correlations, rather than
the full three-dimensional problem. The experiments thus determine a
slice of the full correlation matrix. In this case we need to
determine the effective energy function of the sliced system. It can
be shown \cite{claire, fslice} that projection from $d$~to
$d-1$~dimensions changes the dispersion law in elastic theory from
$\bfk^2$ to~$|\bfk_{\perp}|$ where $\bfk_{\perp}$~is a wave vector in
the projected space and $\bfk$~is the wave vector in the starting
space.

\section{Corruption of correlations by truncation}

\begin{figure}[tb]
  \begin{center}
    \includegraphics{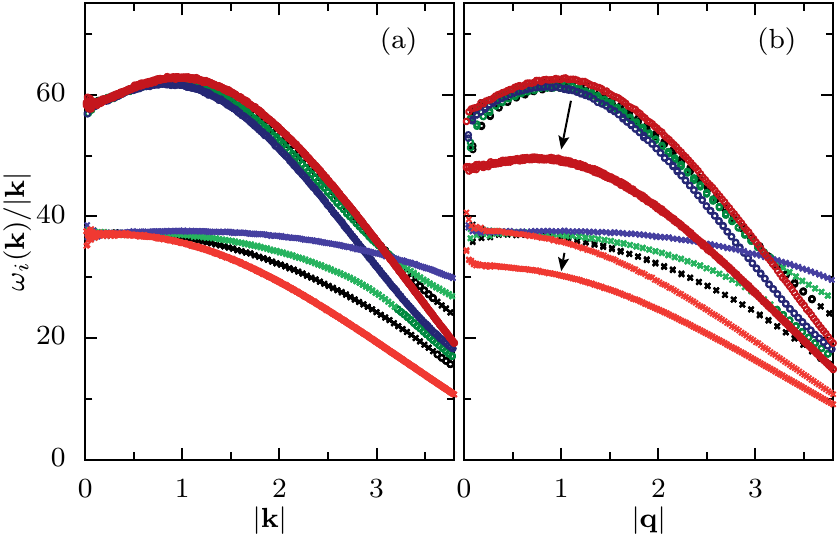}%
    \caption{Windowing artefacts in two-dimensional data.
    (a)~Molecular dynamics data for a hexagonal crystal, analysed for
    its mode structure in periodic boundary conditions. For small wave
    vectors~$k$ we see just two branches corresponding to longitudinal
    (upper, dark) and transverse (lower, light) modes. (b)~The data
    is extracted from an observation window, half the system size in
    both directions, analysed in the same way, plotted are
    $\omega_i(\bfq)/|\bfq|$. The modes split and symmetry-related
    directions are no-longer equivalent. Note windows have non-square
    shape adapted to the underlying hexagonal lattice}%
    \label{fig:two}
  \end{center}
\end{figure}

To demonstrate the problem of working with only the diagonal elements
$C_{ij}(\bfq,-\bfq)$ in truncated data (method~\ref{meth:diag} in the
introduction) we here present results of molecular dynamics
simulations performed in two dimensions on a hexagonal crystal of hard
spheres using event driven methods \cite{rapaport}. In
Fig.~\ref{fig:two}a, we have analysed the mode structure of the whole,
periodic, system. As noted above the long-wavelength mode structure is
described by two elastic constants, $\lambda$, $\mu$ and there is
rotational invariance in long-wavelength correlations. In the figure
this results in there being just two independent intersects for small
wave vectors when we plot~$\omega/k$ as a function of $k$. Here
$\omega^2(\bfk)$ is defined as the eigenvalues of the $2\times 2$
matrix $G_{ij}(\bfk)$ for a given vector~$\bfk$. For the full
simulation volume we conclude that we are able to extract the
effective elastic properties from the diagonal values
$C_{ij}(\bfk,-\bfk)$. Eq.~\eqref{eq:equiv} therefore holds.

However, observation of the data in a finite observation window leads
to considerable modifications in the result, Fig.~\ref{fig:two}b.
We observe a breakdown in rotational invariance, as shown by
the splitting of previously degenerate modes at small wave vectors. In
addition the situation does not improve on increasing the system
size. The main point of the first part of the present paper is
understanding the result of Fig.~\ref{fig:two}b analytically and
finding analytic and numerical methods which allow one to restore the
correct symmetries to the data, in order to correctly characterize
experimental systems and simulation data.

\begin{figure}[tb]
  \begin{center}
    \includegraphics{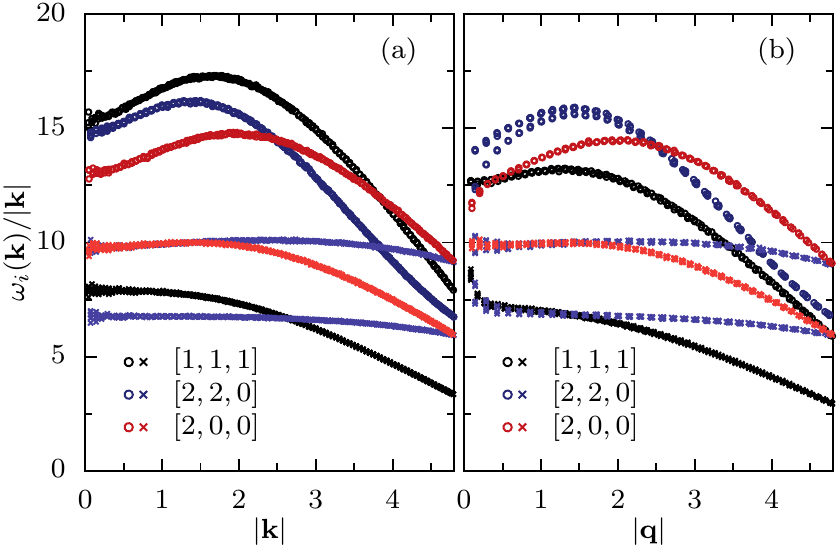}%
    \caption{As for Fig.~\ref{fig:two} but for three-dimensional
    analysis. While the modification of the mode structure is less
    drastic than in two dimensions the longitudinal structure is
    significantly different in the original and windowed data sets.}%
    \label{fig:three}
  \end{center}
\end{figure}

In Fig.~\ref{fig:three} we have performed similar analysis of a
three-dimensional simulation. The results in panel~(a) are more
complicated than in two dimensions because of the influence of cubic
anisotropy. But one clearly sees that the very highest mode in
panel~(a) is displaced to lower~$\omega$ in panel~(b), leading to
important modifications in the effective elastic properties that one
would deduce from the data.

We must conclude that truncating data in a larger experimental system
does not lead to a satisfactory method for measuring elastic
properties. We now demonstrate analytically the origin of the problem.

\subsection{Windowing theory}
\label{sec:windowing}

Let us window the data of an experiment or simulation to a region~$w$
and consider the general two-wave\-vector transform
\begin{equation}
  G_w(\bfq,\bfq') = \int_w \int_w e^{-i(\bfq\dotp\bfr + \bfq'\dotp\bfr')} G(\bfr,\bfr') \;d\bfr\;d\bfr',
\end{equation}
where the integrals are over the observation window. We generalize by
considering a weighting function $W(\bfr)$ which is an arbitrary
positive function. We will always normalize this function such that
\begin{equation}
  \label{eq:normw}
  \int W^2(\bfr) \;d\bfr = \int |W(\bfk)|^2\; \frac{d\bfk}{(2 \pi)^d}=1.
\end{equation}
We find in Fourier space that
\begin{equation}
  \label{eq:results}
  G_w(\bfq,\bfq') := \int W(\bfq-\bfk) W (\bfk
  +\bfq') G(\bfk) \;   \frac{d\bfk}{(2\pi)^d}.
\end{equation}
Windowing to a cube corresponds to a constant $W(\bfr)$ within the
observation zone so that $W(\bfk)$ can be expressed in terms of the
$\sinc$ function, well known from the theory of diffraction.
\begin{equation}
  \label{eq:sinc}
  W(\bfk) = L^{d/2} \prod_{i=1}^d \sinc\Bigl(\frac{k_i L}{2}\Bigr),
\end{equation}
where the product is over the number of dimensions. This type of
window will be analysed in detail below.

We would hope that in the limit of large windows the matrix $G_w(\bfq,
-\bfq')$ is dominated by its diagonal elements and resembles \emph{in
some sense} $G(\bfk)$, thus $W(\bfk)$ should act as a
$\delta$-function. We now examine the conditions in which this
happens. To understand the failures of Fig.~\ref{fig:two}b let us look
at an estimate for the diagonal elements $G_w(\bfq, -\bfq)$. We
consider the scalar theory with $G=1/\scoeff k^2$.  Then
\begin{equation}
  \label{eq:weight}
  G_w(\bfq, -\bfq) = \int |W(\bfk- \bfq)|^2
  \frac{1}{\scoeff k^2}\;\frac{d\bfk}{(2\pi)^d}.
\end{equation}
$|W(\bfk)|^2 $ is simply the Fraunhofer diffraction intensity of the
measurement window.\footnote{The cosine or sine transforms of the
correlations can similarly be expressed in terms the combinations
$\left [G_w(\bfq, -\bfq) \pm G_w(\bfq,\bfq) \right]$.}

\subsubsection{Gaussian window}

Explicit progress can be made using (idealized) Gaussian windows with
\begin{align}
  W(\bfr) &= \frac{1}{(\pi \sigma^2)^{d/4}} e^ {-\bfr^2/2\sigma^2},\\
  W(\bfk) &= (4 \sigma^2 \pi) ^{d/4} e^{-\sigma^2 \bfk^2/2}.
\end{align}
where $\sigma$ is a measure of the real-space width of the window.
Potentially large contributions to the integral in
eq.~\eqref{eq:weight} can come from the central peak of
$W(\bfk{-}\bfq)$ for $\sigma|\bfk-\bfq|=\mathcal{O}(1)$, or from the
divergence of $G(\bfk)$ at~$k=\bfzero$. However the very rapid decay of
$W(\bfk)$ implies that the contribution of the integral from
$k=\bfzero$ is
very small if $\sigma q \gg 1$. We perform the integral
in eq.~\eqref{eq:weight} with a Gaussian window and find
\begin{equation}
  \label{eq:erfi}
  G_w(\bfq,-\bfq) = e^{-q^2 \sigma^2}
  \frac{\sqrt{\pi}\sigma}{\scoeff q}
  \erfi(\sigma q),
\end{equation}
With $\erfi$ the imaginary error function. When $z$~is small,
$\erfi(z) \approx  2z/\sqrt{\pi}$, implying that
\begin{equation}
  G_w(\bfq,-\bfq)= \frac{2\sigma^2}{\scoeff}  \quad \text{for $\sigma q$ small.}
\end{equation}
When $z$ is large $e^{-z^2} \erfi(z)\approx 1/\sqrt{\pi} (z^{-1}+
z^{-3}/2 )$, so that
\begin{equation}
G_w(\bfq,-\bfq) = \frac{1}{\scoeff q^2} \left [ 1 +\frac{1}{2 (\sigma
    q)^2} + \dots \right]\quad \text{for $\sigma q$ large.}
\label{eq:diag}
\end{equation}
Thus with this well-behaved windowing function $G_w(\bfq,-\bfq)$ does
indeed converge to the desired limit, $G(\bfq)$, where corrections are
higher order in $1/\sigma q$.
While for very small values of $\sigma q$ the Gaussian window leads to
an overestimate of the elastic modulus, for most reasonable values of
the parameter we expect a small underestimate in the elastic modulus.
The theory of windowing with a tensorial Green function is given in
Appendix~\ref{sec:elast_gauss} where we show  that we obtain the
correct answer for the both the longitudinal and transverse modes in an
isotropic medium when $\sigma q$ large.

\subsubsection{Discontinuous windows}
\label{sec:discont}

\begin{figure}[tb]%
  \centering
  \includegraphics{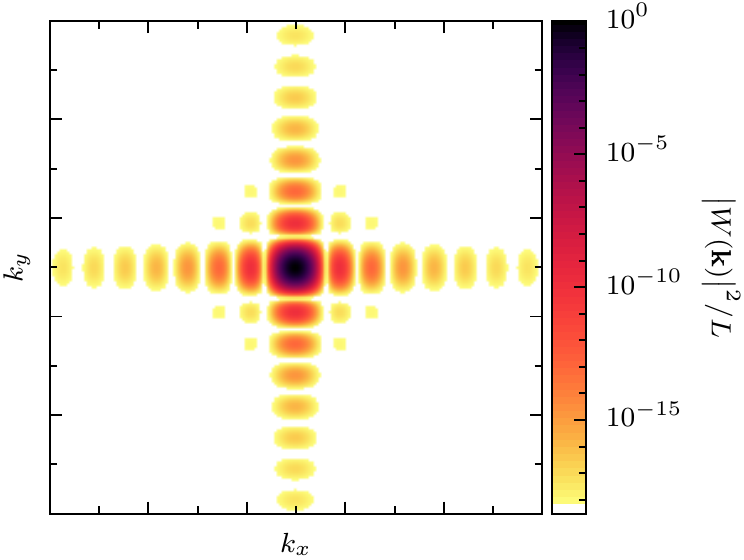}%
  \caption{Fourier transform, $|W(\bfk)|^2$, of the weighting function
  for a square sampling window, eq.~\eqref{eq:sinc}. Perpendicular to
  each side of the window the asymptotic decay of the function is
  slow, $1/k^2$. The decay in general directions is faster, $1/k^4$.
  The slow decay along the symmetry directions leads to major
  artefacts in the reconstructed spectrum.}%
  \label{fig:diffract}
\end{figure}

We now return to the windowing function eq.~\eqref{eq:sinc} which
corresponds to truncating the data. We plot the diffraction intensity
for a two-dimensional square in Fig.~\ref{fig:diffract}. One sees that
the diffraction pattern is characterized by a notable
``cross''-structure in the directions $(1,0)$ and~$(0,1)$. In these
two directions the envelope of Fourier coefficients decays slowly as
$1/k^2$, while off axis the decay is faster, $1/k^4$. We now show that
the decay in $1/k^2$ leads to convergence of the integral in
eq.~\eqref{eq:weight} to an incorrect value and is the origin of the
breakdown in rotational invariance in Fig.~\ref{fig:two}. The
splitting that occurs in Fig.~\ref{fig:two} indeed corresponds to
corruption of modes which are parallel to the slowly decaying
directions in the diffraction pattern of the truncating box. A third
mode in the figure, which is equivalent by symmetry of the hexagonal
lattice but not of the windowing function, remains unaffected by truncation. Note
that such high symmetry directions are those that are the most natural
to analyse in an experimental setup. The problem is that this
diffraction pattern with envelope $1/k^2$ is not sufficiently
``close'' to a $\delta$-function. Qualitatively we see that we are
trying to extract a signal with a power spectrum in $1/k^2$ with a
discontinuous function which has the same power spectrum.

We now argue quantitatively by estimating the integral
eq.~\eqref{eq:weight} with the weighting function of
Fig.~\ref{fig:diffract}. Again two potentially large contributions
come in the neighbourhood of $\bfk=\bfzero$ and $\bfk=\bfq$. For the
integral near $\bfk=\bfq$, the normalization of~$W$ is designed so
that the contribution is exactly that corresponding to the physical
values of~$G$. It is the ball around $\bfk=\bfzero$ that is particularly
problematic, it gives a second, non-negligible contribution to the
integral. With the natural choice of wave vectors $q_i = 2 \pi n/L$
the pole at $\bfk=\bfzero$ is placed on a zero of the diffraction pattern,
however the contribution of a ball of size $\Delta k \sim 1/L$
includes the nearby maxima of the diffraction pattern. The envelope of $|W(\bfk{-}\bfq)|^2$
is slowly varying and can be replaced by a typical value $L^d/(qL)^2$.
In the $x$-direction we have $|W(\bfk{-}\bfq)|^2 \sim |W(\bfq)|^2
(k_xL)^2 $ and find the contribution the contribution of the ball,
\begin{equation}
  |W(\bfq)|^2 \int\limits^{1/L}_{0} (k_xL)^2 \frac{ k^{d-1} dk}{\scoeff k^2} =
  \frac{L^{d-2}}{\scoeff q^2}
  \int\limits^{1/L}_{0} \frac{ k^{d-1} dk}{k^2} \sim \frac{1}{\scoeff q^2}.
\end{equation}
This contribution adds to the $\mathcal{O}(1/\scoeff q^2)$
contribution near $\bfk=\bfq$. The $\sinc$ window over-estimates~$G$,
and thus naturally underestimates effective elastic moduli. In
Fig.~\ref{fig:two}b the dispersion curves which are perpendicular to
the window limits are indeed lower than the correct values.

\subsection{Improved weighting functions}
\begin{figure}[tb]
  \begin{center}
    \includegraphics{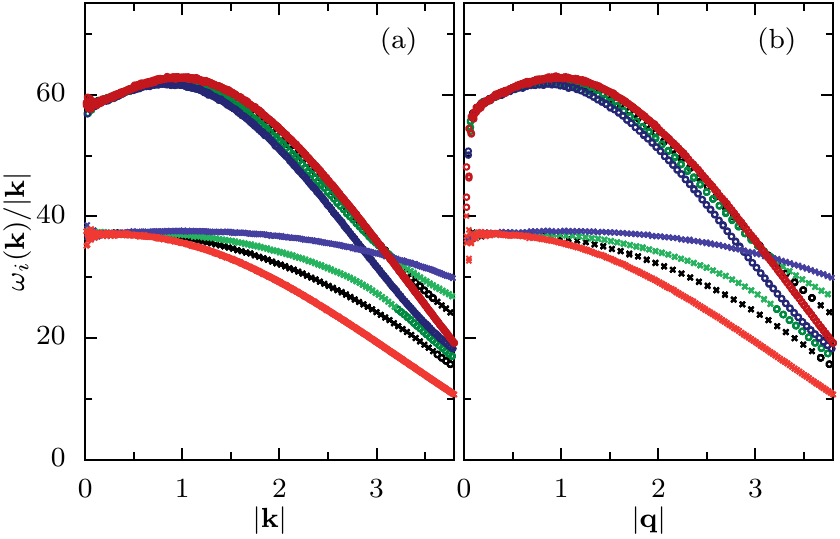}%
    \caption{As Fig.~\ref{fig:two} but using a Hann window in~(b). The
    windowing artefacts are largely eliminated, but the longest
    wavelength modes are modified due to mode mixing, See Appendix A. Skew window
    adapted to simulation box.}%
    \label{fig:twob}
  \end{center}
\end{figure}
\begin{figure}[tb]
  \begin{center}
    \includegraphics{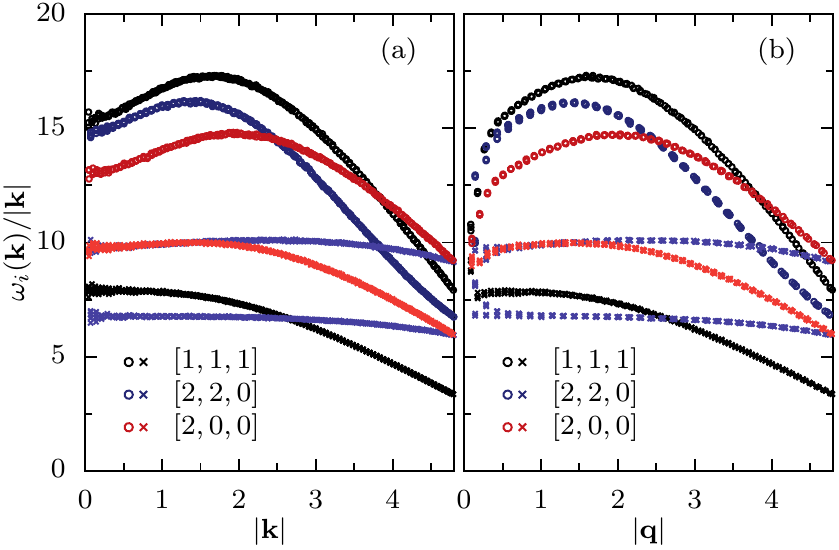}%
    \caption{As Fig.~\ref{fig:three} but using a Hann window in~(b).
    The windowing artefacts are largely eliminated, but the longest
    wavelength modes are modified due to mode mixing, compare
    Appendix~\ref{sec:elast_gauss}. Skew window adapted to simulation
    box.}%
    \label{fig:threeb}%
  \end{center}
\end{figure}
A detailed study of the spectrum properties of windows in signal
processing is given by Nuttall \cite{window}. We see that the
(unrealisable) Gaussian window has excellent properties due to the
suppression of the contribution of the integral at $\bfk=\bfzero$. Sinc-like
windows are inadequate due to their slow decay in Fourier space. We
now show that a product of Hann windows
\begin{equation}
  \label{eq:Hann}
  W(x) =  (1+ \cos {2\pi x/L}),  \quad -L/2 \le x \le L/2,
\end{equation}
has superior Fourier properties. We find
\begin{multline}
  W(k) \sim 2\: \sinc(k L/2) +{}\\ \sinc(kL/2-\pi) + \sinc(kL/2+\pi),
\end{multline}
for which at large wave vectors $|W(k)|^2 \sim 1/k^6$. Such a rapidly
decaying function behaves as a $\delta$-function when tested against a
Green function in $1/k^2$.

We reanalysed our two-dimensional data in Fig.~\ref{fig:twob} using
this Hann window and find greatly improved results, including
restoring of rotational invariance in the data. If one repeats the
analysis of contributions to~$G_w$ coming from~$k=0$ one sees that
they are now sub-dominant. One can trust the results to reconstruct
the true dispersion law.

The corresponding results in three dimension are given in
Fig.~\ref{fig:threeb}. We note that results for the longitudinal modes
drop sharply for small $q$, while the transverse modes rise.
We give a theory of this effect in
Appendix~\ref{sec:elast_gauss} where we
analyse the problem of Gaussian windowing in three dimensional
isotropic media. We show that it is due to a mixing of longitudinal and
transverse modes which occurs for small wavevectors and corrupts
the dispersion relation. Note this is an effect which only occurs for
a few low-lying modes- in distinction to the problems with sharp
windows which lead to all modes in certain directions being corrupted.

\subsection{Observation of slices and projection}
\begin{figure}[ht]
  \centering
  \includegraphics{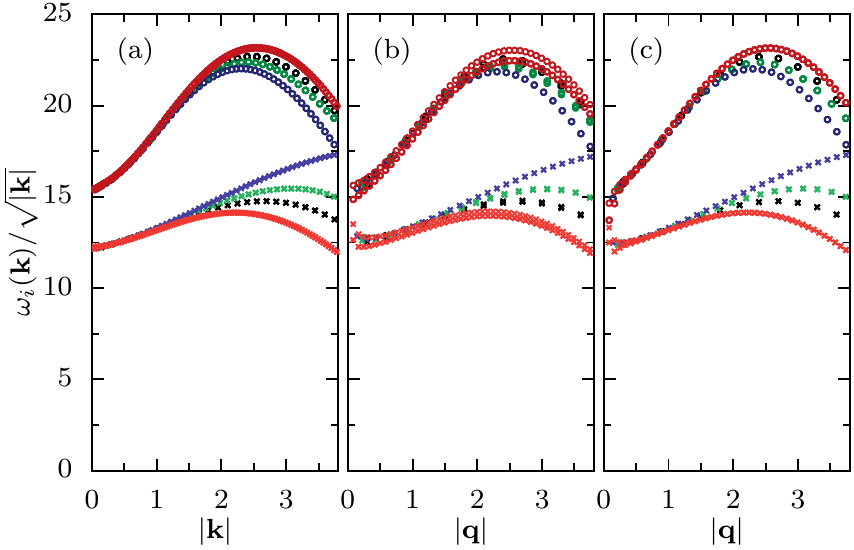}%
  \caption{Projected dispersion curves from the simulation in
  Fig.~\ref{fig:three}. The projected system is two-dimensional and
  isotropic. (a) full system, (b) truncation window, (c) window with Hann
  weighting functions, see eq.~\eqref{eq:Hann}.}%
  \label{fig:proj3D}%
\end{figure}
What is the situation in projection geometries where the effective
dispersion law is $G(\bfk)\sim 1/|k|$, cf.~Refs.\citenum{claire,
fslice}? Here the use of a $\sinc$-function indeed reproduces the
wanted leading term in $1/|q|$ with the correct prefactor, the
integral near $\bfk=\bfq$ is however less divergent than in the case
treated above, so that the origin contributes a sub-leading correction
in $1/q^2$.
At least asymptotically the correct dispersion relation is observed in
$G_w(\bfq,-\bfq)$, though with a leading correction which can no-doubt
be improved with the use of a windowing function which falls to zero
at the edge of the observation zone. We did indeed simulations to test
this point and found the correct reconstructed dispersion law,
Fig.~\ref{fig:proj3D}.

\subsection{Off-diagonal elements}
For the ideal case of the Gaussian window one can estimate the
off-diagonal equivalent to eq.~\eqref{eq:diag}. Via a saddle point
calculation we find
\begin{equation}
  G_w(\bfq, -\bfq') \approx \frac{4}{(\bfq+\bfq')^2} e^{-\sigma^2 (\bfq -\bfq')^2/4}.
\end{equation}
Higher order corrections can also be calculated. An interesting, but
difficult, problem would be to solve for the eigenvalue structure of
this effective matrix.

\section{Mode structure of the full correlation matrix}

We now turn to study of the full correlation matrix $C_{ij}(\bfr,
\bfr')$ of a truncated system, method~\ref{meth:direct} in the
introduction. An experimentalist first measures a set of correlation
functions and assembles them into such a matrix. It is then natural
(and easy with tools such as Matlab) to study the eigenvectors of this
matrix. How are the eigenvectors in a truncated system related to
those of the underlying physical system which is described in terms of
elastic properties? We wish to relate the eigenvalues and eigenvectors
of this matrix in the limit of small wave vectors to those from a true
continuum theory in unbounded space. The particular questions that we
study include the effective boundary conditions for eigenfunctions
that arise from the truncation. We will make particular use of the
scalar analogue of elasticity to simplify the analytic calculations
and to display relevant features of the mathematical problem. We find
the exact analytic solution to the scalar problem in spherical and
circular geometries but use numerical methods in square geometries. We
find the quantisation conditions by direct study of the duplicating
property of an integral operator, and we show that simple Neumann and
Dirichlet boundary conditions do not give the correct mode structure.

\subsection{Scalar elasticity}%
\label{sec:scalar}%
\begin{figure*}[tb]
  \centering
  \includegraphics{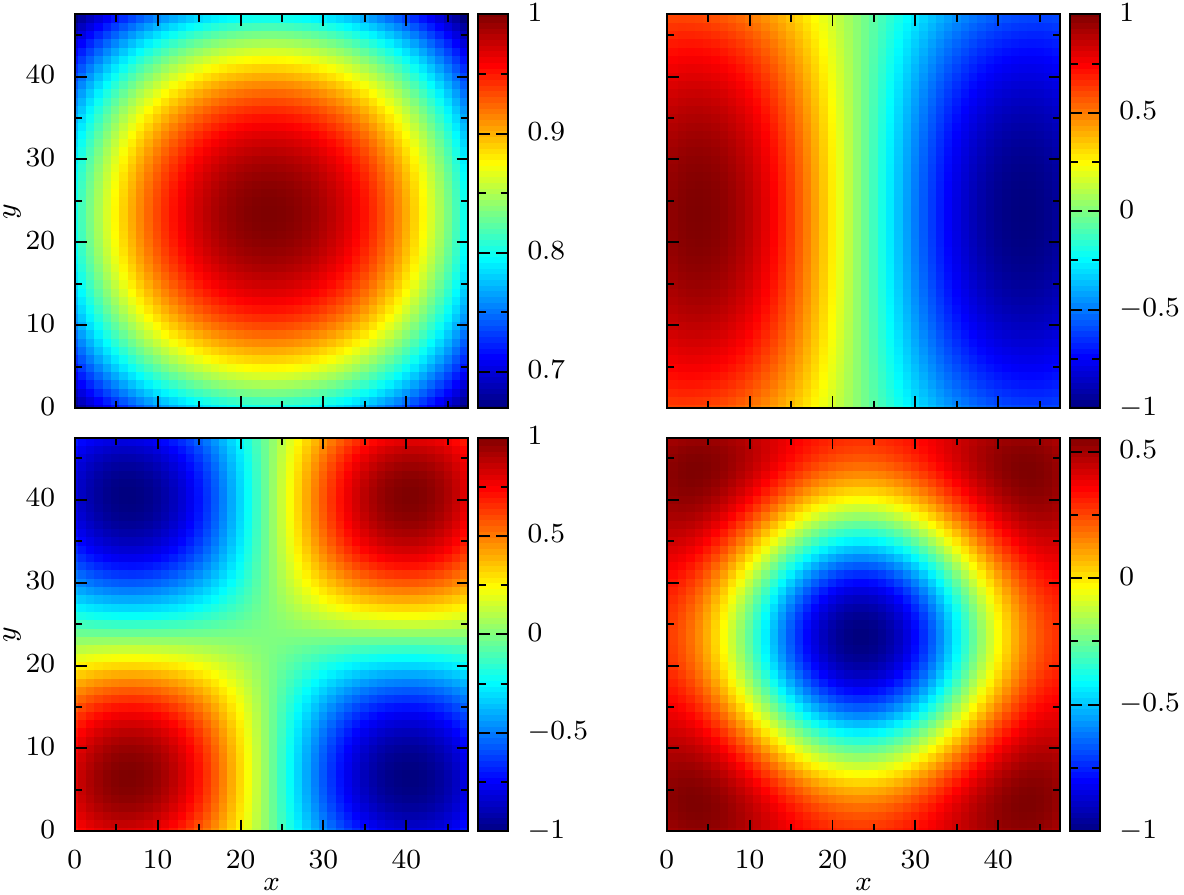}%
  \caption{Amplitudes of modes found in a square observation window
  $48\times 48$ pixels, using scalar elasticity. Top left to bottom
  right decreasing eigenvalues of the integral operator. Note that
  first mode is not constant within the observation volume despite it
  being the most uniform mode. The second mode displays a noticeable
  non-sinusoidal form.}%
  \label{fig:modes}%
\end{figure*}%

We have noted that the problem of scalar elasticity is linked to the
properties of the Laplacian operator which corresponds to the
Helmholtz eigenvalue equation:
\begin{equation}
  \label{eq:helmholtz}
  [\nabla^2 + k^2] \psi=0,
\end{equation}
The correlation matrix from scalar elasticity is then the form
$C(\bfr, \bfr') =1/4 \pi |\bfr -\bfr' | $.

If we construct a (continuum) matrix from the correlations we are thus
interested in the following 
integral equation
\begin{equation}
  \label{eq:integral}
  \int_V \frac{1}{4\pi|\bfr-\bfr'|} \psi(\bfr) \; d^3\bfr = \Lambda \psi(\bfr'),
\end{equation}
The eigenvalue is $\Lambda$, and we are
restricted to a finite volume~$V\subset\Reals^3$.  The mathematical
difficulty comes from the arbitrary choice for the shape of~$V$.

Consider equation eq.~\eqref{eq:integral} with $\bfr'$ within the
observation volume then we can act on this equation with $-\nabla^2$
acting on the $\bfr'$ coordinate to find
\begin{equation}
  \int_V \delta(\bfr-\bfr') \psi(\bfr)\;d^3\bfr
  = \psi(\bfr')
  = -\Lambda \nabla^2 \psi(\bfr'),
\end{equation}
which is the Helmholtz equation with eigenvalue $\Lambda= k^{-2}$.
Thus the matrix of correlation functions has eigenvectors~$\psi_k$
which are closely related to those of the corresponding differential
equation, however we will now show that the boundary conditions are
different.
We have the equations
\begin{gather}
  -\nabla^2  G(\bfr-\bfr') = \delta(\bfr-\bfr'),  \label{eq:1}\\
  (\nabla^2 +k^2) \psi_k(\bfr) = 0. \label{eq:2}
\end{gather}
Multiplying eq.~\eqref{eq:1} by $\psi_k(\bfr)$ and eq.~\eqref{eq:2} by
$G(\bfr-\bfr')$ leads to
\begin{equation}
  \int_V\; d\bfr\; \left [ \psi_k \nabla^2 G - G \nabla^2 \psi_k  - k^2
    \psi_k G \right ] = - \psi_k(\bfr').
\end{equation}
Using Green's second identity we find
\begin{equation*}
  \oint_{\partial V} d{\bf S}_r \dotp \left[\psi_k\nabla G - G\nabla\psi_k\right]
  =  -\psi_k(\bfr') + k^2 \int_V d\bfr \psi_k(\bfr)G(\bfr,\bfr').
\end{equation*}
We recognize the right-hand side as the eigenvalue equation in
integral form, thus the condition
\begin{equation}
  \oint_{\partial V} d{\bf S}_r\dotp\left[\psi_k(\bfr)\nabla G(\bfr,\bfr')
  - G(\bfr,\bfr')\nabla\psi_k(\bfr) \right] = 0 \quad \forall \bfr'
\end{equation}
is required for $\psi_k$~being an eigenmode both of the elastic
problem~\eqref{eq:helmholtz} and of the truncated integral
equation~\eqref{eq:integral}. For any eigenvector $\psi_k$, this is a
set of integral conditions, true for each interior point $\bfr'$. In
spherically symmetric domains~$V$, see Sec.~\ref{sec:symmsol} below,
they reduce to boundary conditions valid  on the surface. In
the general case, this cannot be assumed to be the case.

\subsection{Tensor elasticity}

While performed in the simplest scalar form, the above theory can be
easily replaced by its tensorial equivalent for an elastic medium. The
only difference is the use of Betti's identity for the stress tensor
rather than Green's second theorem\cite{betti}.
Let
\begin{align}
  \label{eq:beg_elast}
  \Delta^*_{ik} &:= C_{ijkl}\partial_j\partial_l \quad\text{in general, and}\\
  \Delta^* &= (\lambda+\mu) \grad\div + \mu \nabla^2\quad\text{if isotropic,}
\end{align}
then
\begin{equation}
  \oint_{\partial V} dS\; \bigl[(\partial_\bfN G)\dotp\bfu - G\dotp(\partial_\bfN\bfu)\bigr]
  = \int_V \bigl[(\Delta^*G)\dotp\bfu - G\dotp(\Delta^*\dotp\bfu)\bigr]\; d\bfr
\end{equation}
with $\partial_\bfN \bfu$ the normal stress and $\partial_\bfN G$ the
normal derivative of the Green function,
\begin{equation}
  \partial_\bfN := C_{ijkl}\partial_j N_l.
\end{equation}

The truncated and the elastic eigenvalue problems then have the same
eigenfunctions if the integral relation
\begin{equation}
  \label{eq:end_elast}
  \oint_{\partial V} dS\; \bigl[(\partial_\bfN G)\dotp\bfu -
  G\dotp(\partial_\bfN\bfu)\bigr] = \bfzero
\end{equation}
holds for any eigenvector~$\bfu$ and for any point~$\bfr'$ (the
integrals and derivatives in
eqs.~\eqref{eq:beg_elast}--\eqref{eq:end_elast} act on~$\bfr$).

\subsection{Scalar spherically symmetric solutions}%
\label{sec:symmsol}

We now derive the exact quantization conditions in rotationally
symmetric geometries, again studying the scalar vibrational problem:
Regular solutions to the Helmholtz equation in spherically symmetric
geometries can be written in terms of spherical Harmonics~$Y^m_l$ and spherical Bessel
functions~$j_l$,
\begin{equation}
  \label{eq:psi}
  \Psi^m_l(\bfr) = j_l(kr)\, Y^m_l (\Omega).
\end{equation}
$\Omega=(\theta, \phi)$ is a solid angle in spherical polar
coordinates. For $l=0$ we note that
\begin{equation}
 \Psi^0_0= j_0(kr) = \frac{\sin(kr)}{kr}.
\end{equation}
Let us show that $\Psi^m_l$ is also a solution to the truncated
integral equation~\eqref{eq:integral} for correct choices of~$k$.

As domain we consider the ball of radius~$R$ around the origin,
$V=B_R(\bfzero)$, so that the integral operator acting on the
trial function is given by
\begin{equation}
  I = \int\limits_{B_R(\bfzero)} \frac{1}{4\pi|\bfr-\bfr'|}
  \Psi^m_l(\bfr)\:  d^3\bfr.
\end{equation}

We use the identity ($r_<={\rm min}(r,r')$ and $r_>={\rm max}(r,r')$)
\begin{equation*}
  \frac{1}{|\bfr -\bfr' |} = \sum_{l=0}^{\infty} \frac{r_<^l}{r_>^l}
  \frac{4 \pi}{2l +1} \sum_{m=-l}^{l} (-1)^m Y^m_l(\Omega) Y^{-m}_l (\Omega')
\end{equation*}
to break the integral into a radial and an angular part, and then use
the fact that after the angular integrals only a single spherical
harmonic survives so that we must evaluate
\begin{equation}
  I=\frac{Y^m_l}{2l +1} \int_0^R \frac{1}{|\bfr- \bfr'|} j_l(kr)
  \frac{r^l_<}{r^{l+1}_>}\;r^2\, dr.
\end{equation}
We explicitly split the integration to find
\begin{equation*}
  \label{eq:transform}
   I = \frac{Y^m_l}{2l +1} \left( \int_0^{r'} \frac{r^{l+2}}{r'^{l+1}}  j_l(kr) \; dr + \int_{r'}^R
  \frac{r'^l}{r^{l-1}} j_l(kr)
  \; dr  \right).
\end{equation*}
We now use the identities \cite{abramowitz}
\begin{align}
  \int z^{n+1} j_{n-1}(z) &= z^{n+1} j_n(z),\\
  \int \frac{1}{z^{n}} j_{n+1}(z) &= -\frac{1}{z^n} j_n(z),\\
  j_{n+1}(z) + j_{n-1}(z)&= \frac{(2 n +1)}{z} j_n(z),
\end{align}
to transform the integrals into
\begin{equation}
  I= \frac{Y^m_l(\Omega')}{k^2} \left( j_l(kr') -
    \frac{j_{l-1}(kR)}{2l+1} \frac{(kr')^l}{(kR)^{l-1}} \right).
\end{equation}
This is of the original form~$\Psi^m_l$ if $j_{l-1}(kR)=0$ which
serves to obtain possible values for~$k$:
\begin{align}
  &l=0:&  &\cos(kR)=0 \qquad\Rightarrow\qquad kR=(n{+}1/2)\pi,\\
  &l=1:&  &\sin(kR)=0 \qquad\Rightarrow\qquad kR=n\pi,
\quad\text{etc.}
\end{align}
The condition $zj_{l-1}(z)=0$ can be reformulated as\cite{abramowitz}
\begin{equation}
  \label{eq:robin}
  z \frac{d}{dz} j_l (z) +(l{+}1)j_l(z)=0, \quad\text{with}\quad z=kR,
\end{equation}
which is a mixed (Robin) boundary condition for~$\Psi^m_l$ on the
sphere. Now we understand how we change the original elastic
problem~\eqref{eq:helmholtz} if we truncate the domain: Truncating is
equivalent to applying the boundary conditions~\eqref{eq:robin} to the
different solutions of eq.~\eqref{eq:helmholtz}. Note that for
different~$l$ we have different boundary conditions.

A very similar expansion in $(r_</r_>)$ can be performed in two dimensions for a
disk, where the eigenfunctions $e^{i m \theta} J_m(kr)$ give rise to
the eigen-equation $J_{m-1}(kR)=0$ for $m>0$. For $m=0$ the
corresponding equation is 
\begin{equation}
 J_0(z) + z \log(z/z_0) J_1(z) =0, \label{eq:logev}
\end{equation} with
$z=kR$ and $z_0$ a reference radius.

\subsection{Square geometry}

\begin{figure*}[ht]
  \centering
  \includegraphics{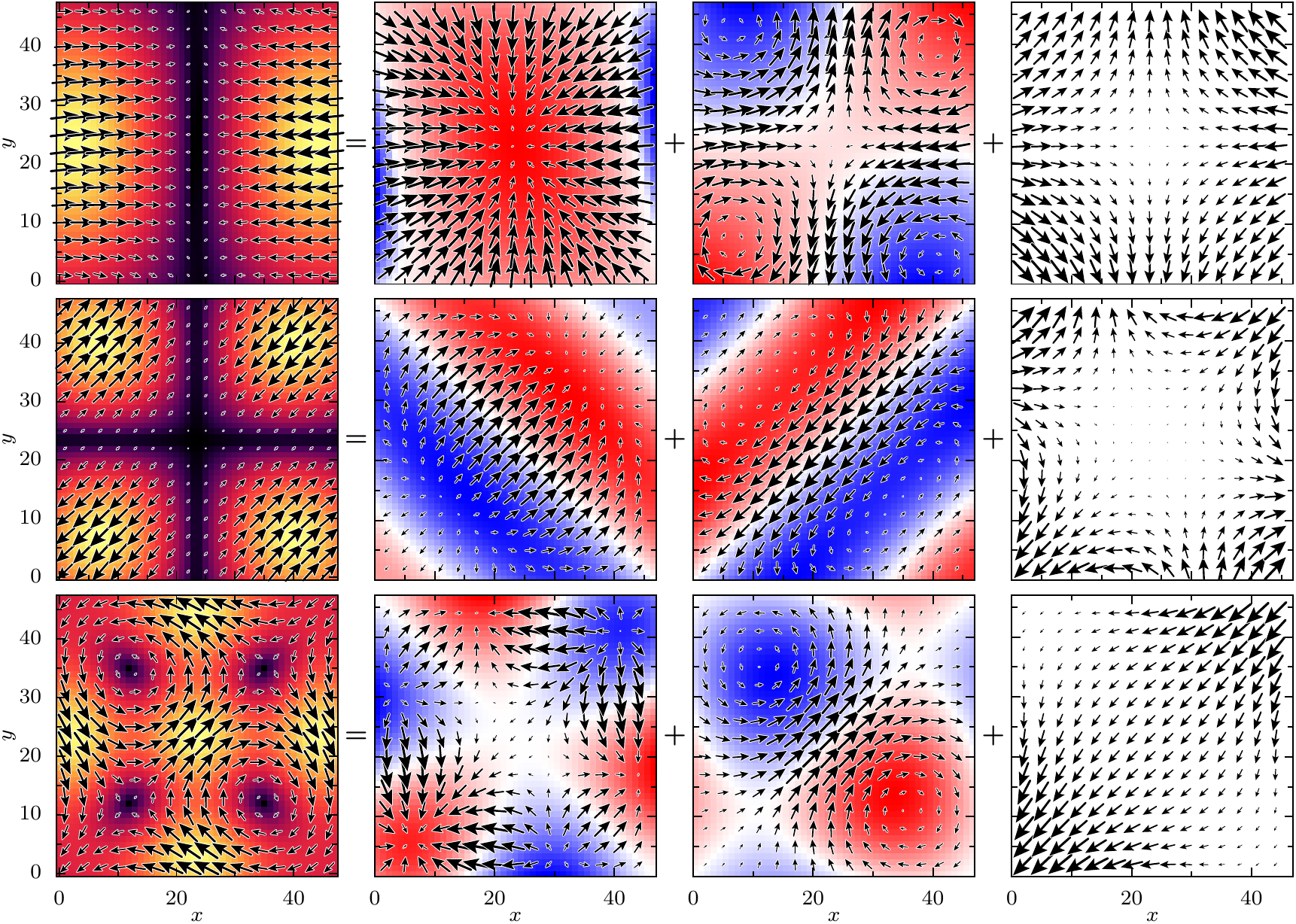}%
  \caption{Three eigenmodes of the truncated system: Left panels show
    the displacement field, colours indicate absolute value
    (black=small, yellow=large). The three columns on the right show
    the Helmholtz--Hodge decomposition into longitudinal, transverse
    and harmonic parts. The blue/red colour shows the divergence
    (second column) and the rotation (third column) of the
    displacement field (red=negative, blue=positive). $\lambda=\mu$.}%
  \label{fig:modes2}%
\end{figure*}

We were unable to solve the integral equation eq.~\eqref{eq:integral}
in the geometry of a square or cubic box. We thus proceed by numerical
investigation. We generate the Green function in a large, periodic box
of dimension $500 \times 500$ using fast Fourier transforms
\cite{fslice}. We then truncate the Green function to a window of
dimension $48 \times 48$. This Green function is then diagonalized
using standard dense algebra packages included in Matlab. The
structure of the modes that is found is demonstrated in Fig.
\ref{fig:modes}. The simplest mode (top left panel) has a largely
constant amplitude over the observation region, it does have a gentle
peak at the centre, however, which is some $30 \%$ higher than the
value of the function in the corners. It is interesting to note that
this mode has finite energy- it does not correspond to a zero mode of
the Helmholtz system. The next mode (top right panel), which is
two-fold degenerate, has the nature of a wave within the box. However
we see that it is definitely non-sinusoidal and the contours of
constant amplitude bow out at the edge of the observation volume.

The methods generalize to vector elasticity in two dimensions. We use
a modification of the method of Ref.~\citenum{fslice} to generate a
discretised version of the matrix~$D$ in eq.~\eqref{eq:christoffel},
with $\nu=0$. In particular we choose a discretised dispersion
relation
\begin{multline}
  D(\bfk) = \mu(4-2\cos k_x-2\cos k_y)
  \begin{pmatrix} 1 & 0 \\ 0 & 1 \end{pmatrix} +\\
  (\lambda+\mu)
  \begin{pmatrix} 2-2\cos k_x & \sin k_x \sin k_y \\ \sin k_x \sin k_y
    & 2-2\cos k_y \end{pmatrix}
\label{eq:new}
\end{multline} 
and use the fast Fourier transform to generate the corresponding
real-space form $G_{ij}(\bfr)$. We truncate this Green function to a
square and diagonalize. The dispersion relation differs from our
previous choice\cite{fslice} and has the advantage of preserving more
properties of the continuum elastic theory that we wish to study. In
particular our previous choice leads to a subdominant contribution to
the $G_{xy}(\Delta x, 0)$.  In our previous work it decays as
$1/(\Delta x)^2$ for large separations $\Delta x$. The new form
eq.~(\ref{eq:new})
gives
zero for this quantity.

In the leftmost column of fig.~\ref{fig:modes2} we plot the vector
displacement fields of three of the lower eigenmodes in an isotropic
elastic medium, windowed to $48 \times 48$ in a system of dimensions
$500\times500$. Both the colour coding in black/yellow and the size of
the arrows indicate the absolute value of the displacements. In these
modes we find similar bow-shaped structures in the modes to those
found in the scalar problem. It is interesting to note a number of
properties of the figures. When we look at the distribution of
amplitudes in the top panel, the maximum occurs at some distance from
the edge of the sample and displays a vertical gradient in colour --
this shows that despite the mode being largely longitudinal in nature
it does display both longitudinal and transverse characters.

The longitudinal and transverse nature of these modes is better
studied in an explicit Helmholtz--Hodge decomposition, which is
displayed in the second to fourth columns of
fig.~\ref{fig:modes2}. This requires some explanation. We want to
split a vector field into a rotation-free part, $\bfnabla\phi$ and a
divergence-free part~$\bfgrot\psi$, with two scalar fields $\phi$ and
$\psi$. The vector differential operator $\bfgrot :=
(-\partial_y, \partial_x)^T$ replaces the $\curl$ in two
dimensions\footnote{$\psi$ can be considered the $z$ component of the
  vector potential.}. The rotation-free part should be
``longitudinal'', and the divergence-free part
``transverse''. However, this decomposition is unique only in infinite
space. 
In the present case with a finite window, a third component~$\bfh$ may
be required, having zero divergence and rotation.  The decomposition
then reads
\begin{equation}
  \label{eq:decomp}
  \bfu = \bfnabla\phi + \bfgrot\psi + \bfh.
\end{equation}
After scalar multiplications with $\bfnabla$ and with $\bfgrot$, one
finds that the scalars $\phi$ and $\psi$ satisfy Poisson equations
with divergence and rotation of the original field as sources,
\begin{equation}
  \label{eq:decomp_poisson}
  \laplace\phi = \bfnabla\dotp\bfu, \qquad
  \laplace\psi = \bfgrot\dotp\bfu.
\end{equation}
We now see that in the finite window the solutions to these equations
are no longer unique but depend also on the boundary conditions we
apply. One may obtain a possible decomposition into divergence-free and
rotation-free parts by solving for~$\phi$ with some boundary conditions
imposed, e.~g.~Dirichlet or Neumann; the rest, $\bfu-\bfnabla\phi$ is
then representable as~$\bfgrot\psi$. If specific boundary conditions
are required also for~$\psi$, then there is a third
contribution~$\bfh$ which obeys the equation
\begin{equation}
  \bfgrot\dotp\bfh = \bfnabla\dotp\bfh  = 0,\quad
  \laplace\bfh=\bfzero.
\end{equation}
In our case we lack a reasonable justification of such a boundary
condition for $\phi$ or $\psi$, they give rise to visually
unreasonable fields- for instance normal fluxes which are forced to
zero at the edge of the box. We rather choose to come as close as
possible to what we associate with ``longitudinal'' and ``transverse''
in the infinite system. We therefore do not impose a boundary
conditions at the box edges but extend eqs.~\eqref{eq:decomp_poisson}
to infinity by padding the right-hand sides with zero and requiring
the solutions to vanish at infinity. The solution to this problem has
already been discussed in sec.~\ref{sec:scalar} and is the convolution
of the right-hand sides of eq.~\eqref{eq:decomp_poisson} with minus
the scalar Green function. The scalar fields $\phi$ and $\psi$ contain
the divergence and the rotation of the original field~$\bfu$, and we
believe that these solutions are the least perturbed by any boundary
effects. The harmonic field $\bfh$ is then simply what remains after
subtraction of $\bfnabla\phi$ and $\bfgrot\psi$.
\begin{table}[tb]
  \centering
  \begin{tabular}{cccc}
    $\int |\bfu|^2$ & $\int|\bfnabla\phi|^2$ & $\int|\bfgrot\psi|^2$ & $\int|\bfh|^2$ \\\hline
    1.00 & 0.42 & 0.04 & 0.25 \\
    1.00 & 0.30 & 0.29 & 0.19 \\
    1.00 & 0.02 & 1.06 & 0.42
  \end{tabular}%
  \caption{L$_2$~norms of the vector fields displayed in
    fig.~\ref{fig:modes2}, in the same order.}%
  \label{tab:modes2}
\end{table}

The three contributions of eq.~\eqref{eq:decomp} are displayed in the
three rightmost columns of fig.~\ref{fig:modes2} (in the same
order). The blue/red colour coding shows the right-hand sides of
eqs.~\eqref{eq:decomp_poisson}, red for negative values and blue for
positive. In the second column, displaying~$\bfnabla\phi$, we see how
the arrows all head to the minima/maxima of the divergence field. In
the second column, displaying~$\bfgrot\psi$, the arrows turn in
positive or negative sense around the minima/maxima. The harmonic
contribution in the third column has zero divergence and zero
rotation. The arrows in all panels are scaled independently to best
visualize the fields. For quantitative comparison we give the
L$_2$~norms in table~\ref{tab:modes2} \footnote{The decomposition is
  not orthogonal so that the sum of the three individual terms is not
  unity}. One sees that the mode in the first row is mainly
longitudinal, but that there is an important harmonic (quadrupolar)
contribution. We found another mode of similar symmetry which was
dominated by its transverse part (not shown). The last line of the
figure shows again a mainly transverse mode, but here the harmonic
field is not quadrupolar but is of lower order than the dominant
transverse part. Most interestingly, the middle row shows a mode where
both longitudinal and transverse contributions are equally
important. This shows clearly that the truncation mixes these two
natures of the modes.

A numerical study of the scaling of the mode energies with size of the
truncating box, $\ell$ confirmed that, as expected, eigenvalues scale
as $\ell^2$; thus $\Lambda/\ell^2$ is the object which contains
information about material properties. However we note that
eq.~(\ref{eq:logev}) implies that in a disk geometry there can be a
slow logarithmic cross-over for certain modes. One might expect
similar logarithmic corrections in the square geometry too.

\section{Conclusions}

Data analysis, with both numerical and experimental data often require
windowing. When correlation data is simply truncated it can lead to
substantial artefacts in the measured amplitudes and can mislead as to
the exact values of elastic constants. These errors can be
considerably reduced by using windowing functions which decay faster
in Fourier space. In particular we found that the Hann window gives
good results.

The experimental analysis of windowed data also gives rise to
interesting questions as to the nature of the observed eigenmodes. We
have shown that the correlation functions give rise to problems which
satisfy an interesting integral condition involving the correlation
functions of the experimental system. We have made a study of the
eigenvalue problem for scalar elasticity and showed how to find exact
analytic solutions of the integral equation in spherical and circular
geometries. In square geometries we exhibited eigenfunctions which
deviate noticeably from plane waves. We note that the use of integral
equations to characterize observed experimental correlations is known
in field such as statistical analysis and atmospheric physics \cite{basis}.

\appendix
\section{Elasticity with Gaussian window}%
\label{sec:elast_gauss}

We present here the
calculation for a tensorial Green function in an isotropic
three-dimen\-sional medium, when analysed using Gaussian windowing.
The longitudinal part of the Green function of the original system can
be written in the form
\begin{equation}
  \label{eq:AB}
  G^{(l)}(\bfk) = \frac{1}{\lambda + 2 \mu} \frac{|\bfk\rangle
  \langle\bfk|}{k^2},
\end{equation}
with unit column vectors~$|\bfk\rangle = \bfk/k$ and their adjoint row
vectors~$\langle\bfk|$. The Gaussian weighting of
eq.~(\ref{eq:weight}) involves only a single external vector quantity
$\bfq$. Thus we can deduce that
\begin{equation}
  \label{eq:Gq}
  (\lambda + 2 \mu)G^{(l)}_w(\bfq, -\bfq) = A(q) {\bf I} + B(q) |\bfq\rangle \langle\bfq|
\end{equation}
for two as yet unknown functions $A$ and $B$. We note that this form
involves both longitudinal and transverse parts in the new
variables. We take the scalar product of eq.~(\ref{eq:AB}) and
eq.~(\ref{eq:Gq}) with $\bfq$ and secondly study the trace of these
equations to deduce that
\begin{align}
  A(q)+ B(q) =\int |W(\bfq-\bfk)|^2 \frac{ (\hat \bfq \dotp \hat
    \bfk)^2 }{k^2}
  \frac{d\bfk}{(2 \pi)^3} \\
  3 A(q)+ B(q)=\int |W(\bfq-\bfk)|^2 \frac{ 1 }{k^2} \frac{d\bfk}{(2
    \pi)^3}
\end{align}
We perform the angular integrals, then recognize the radial integrals
as being related to imaginary error functions:
\begin{align}
  A(q)+B(q) &= \frac{1}{q^2} - e^{-\sigma^2 q^2} \frac{ \sqrt{\pi}
    \erfi(\sigma q)}{2\sigma q^3}\\
  3A(q) +B(q) &= e^{-q^2 \sigma^2} \frac{\sqrt{\pi}\sigma}{q}
  \erfi(\sigma q)
\end{align}
The transverse part of the Green function
\begin{equation}
  G^{(t)}_{ij} (\bfk) = \frac{{\bf I}-|\bfk\rangle \langle\bfk|}{\mu k^2}
\end{equation}
also gives a contribution which can be expressed in terms of the
functions $A$ and $B$:
\begin{equation}
  \label{eq:Gt}
  \mu G^{(t)}_w(\bfq, -\bfq) = (2A(q) + B(q)) {\bf I} - B |\bfq \rangle
  \langle \bfq|.
\end{equation}
The full Green function is then the sum of the contributions of
eq.~\eqref{eq:Gq} and eq.~\eqref{eq:Gt}.

\begin{figure}[htb]
  \centering
  \includegraphics{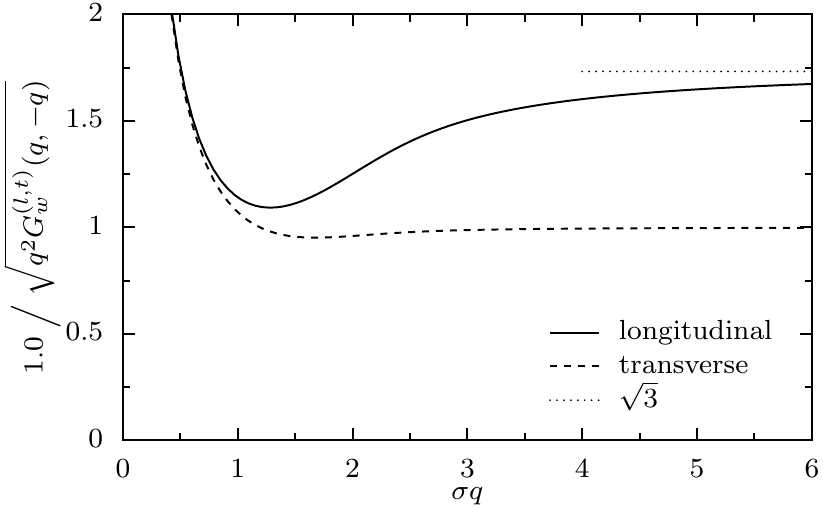}%
  \caption{Effective $\omega/q$ as a function of $\sigma q$ for a
  three dimensional system with $\lambda/\mu=1$, $\nu=0$. For large
  values of $\sigma q$ the ratio of the two curves converges to
  $\sqrt{3} $. However mixing of the modes leads to a strong drop in
  the estimate of the longitudinal stiffness for $\sigma q\approx1$.}%
  \label{fig:erfi}%
\end{figure}

When $\sigma q$ is large $A \approx1/(\sigma^2q^4)$, whereas
$B(q)\approx 1/q^2$. Thus the reconstruction does not mix the
longitudinal and transverse components of the response which are both
found correctly. For intermediate values of $\sigma q$, where $A$
cannot be neglected, the transverse
and longitudinal modes do mix, in a manner similar that we found with
the small-wavevector reconstructions using the Hann window,
Fig.~{\ref{fig:threeb}}. In particular the longitudinal stiffness is
strongly underestimated. As a specific example we take $\lambda/\mu=1$
and plot in Fig~\ref{fig:erfi} the two effective values of $\omega$ as
a function of $\sigma q$.

Matching the mean squared width of a Hann window to a Gaussian gives
$\sigma^2= L^2(1/12 - 1/(2\pi^2) )$, giving an approximate relation
between our analytic calculations on Gaussian functions and practical
windows in the experimental situation. For the first mode in a square
sample for which $q=2 \pi/L$ we find $\sigma q \approx 1.1$.

\footnotesize{
\providecommand*{\mcitethebibliography}{\thebibliography}
\csname @ifundefined\endcsname{endmcitethebibliography}
{\let\endmcitethebibliography\endthebibliography}{}

}
\end{document}